\documentclass[11pt]{article}

\usepackage{kotex}
\usepackage{titlesec}
\usepackage{bookmark}
\usepackage{import}
\usepackage[utf8]{inputenc}
\usepackage[a4paper]{geometry}
\usepackage{caption}
\usepackage{palatino}

\usepackage[nodisplayskipstretch]{setspace}
\usepackage{indentfirst}
\usepackage{graphicx}
\graphicspath{{./images}}

\usepackage{hyperref}
\hypersetup{
    colorlinks=true,
    linkcolor=red,
    unicode=true,
    urlcolor=black,
    citecolor=red,
    citecolor=blue,
    linkcolor=blue,
}

\usepackage[
    backend=biber,
    style=numeric,
    bibstyle=numeric-verb,
    sorting=none
]{biblatex}
\usepackage{cleveref}

\usepackage[straightquotes]{newtxtt}
\renewcommand{\tt}[1]{\texttt{#1}}

\crefname{lstlisting}{listing}{listings}
\Crefname{lstlisting}{Listing}{Listings}

\usepackage{listings}
\usepackage{xcolor}
\colorlet{punct}{red!60!black}
\definecolor{delim}{RGB}{20,105,176}
\colorlet{numb}{magenta!60!black}

\lstdefinelanguage{json}{
    showspaces=false,
    showtabs=false,
    breaklines=true,
    postbreak=\raisebox{0ex}[0ex][0ex]{\ensuremath{\color{gray}\hookrightarrow\space}},
    breakatwhitespace=true,
    basicstyle=\ttfamily\footnotesize,
    upquote=true,
    morestring=[b]",
    literate=
    *{0}{{{\color{numb}0}}}{1}
     {1}{{{\color{numb}1}}}{1}
     {2}{{{\color{numb}2}}}{1}
     {3}{{{\color{numb}3}}}{1}
     {4}{{{\color{numb}4}}}{1}
     {5}{{{\color{numb}5}}}{1}
     {6}{{{\color{numb}6}}}{1}
     {7}{{{\color{numb}7}}}{1}
     {8}{{{\color{numb}8}}}{1}
     {9}{{{\color{numb}9}}}{1}
     {:}{{{\color{punct}{:}}}}{1}
     {,}{{{\color{punct}{,}}}}{1}
     {\{}{{{\color{delim}{\{}}}}{1}
     {\}}{{{\color{delim}{\}}}}}{1}
     {[}{{{\color{delim}{[}}}}{1}
     {]}{{{\color{delim}{]}}}}{1},
}

\begin{document}

\begin{center}

    \vspace*{50pt}

    {\fontsize{20}{0}\selectfont Secure IAM on AWS with Multi-Account Strategy}

    \vspace*{12pt}

    {\fontsize{18}{0}\selectfont (다중 계정을 이용한 안전한 AWS 권한 관리)}

    \vspace*{80pt}

    {\fontsize{18}{0}\selectfont 지도교수 : 허충길}

    \vspace*{80pt}

    {\fontsize{20}{0}\selectfont 이 논문을 공학학사 학위 논문으로 제출함.}

    \vspace*{85pt}

    {\fontsize{18}{0}\selectfont 2023년 \hspace*{8pt} 12월 \hspace*{8pt} 26일}

    \vspace*{70pt}

    {\fontsize{20}{0}\selectfont 서 울 대 학 교 \hspace*{5pt} 공 과 대 학}

    \vspace*{10pt}

    {\fontsize{20}{0}\selectfont 컴 ~퓨 ~터 ~공 ~학 ~부}

    \vspace*{10pt}

    {\fontsize{20}{0}\selectfont 이 ~ 성 ~ 찬}

    \vspace*{40pt}

    {\fontsize{20}{0}\selectfont 2024년 \hspace*{5pt} 2월}
\end{center}

\thispagestyle{empty}
\pagebreak

\vspace*{8pt}

\begin{center}
    {\fontsize{16}{0}\selectfont \textbf{Abstract}}

    \vspace*{15pt}

    {\fontsize{20}{20}\selectfont {Secure IAM on AWS with Multi-Account Strategy}}
\end{center}

\vspace*{20pt}

\begin{flushright}
    {
        \fontsize{14}{0}\selectfont

        Sungchan Yi

        \vspace*{10pt}

        Department of Computer Science and Engineering

        \vspace*{10pt}

        Seoul National University

    }
\end{flushright}

\vspace*{10pt}

{\fontsize{11}{13}\selectfont

Many recent IT companies use cloud services for deploying their products, mainly because of their convenience. As such, cloud assets have become a new attack surface, and the concept of cloud security has emerged. However, cloud security is not emphasized enough compared to on-premise security, resulting in many insecure cloud architectures. In particular, small organizations often don't have enough human resources to design a secure architecture, leaving them vulnerable to cloud security breaches.

We suggest the multi-account strategy for securing the cloud architecture. This strategy cost-effectively improves security by separating assets and reducing management overheads on the cloud infrastructure. When implemented, it automatically provides access restriction within the boundary of an account and eliminates redundancies in policy management. Since access control is a critical objective for constructing secure architectures, this practical method successfully enhances security even in small companies.

In this paper, we analyze the benefits of multi-accounts compared to single accounts and explain how to deploy multiple accounts effortlessly using the services provided by AWS. Then, we present possible design choices for multi-account structures with a concrete example. Finally, we illustrate two techniques for operational excellence on multi-account structures. We take an incremental approach to secure policy management with the principle of least privilege and introduce methods for auditing multiple accounts.

}

\vspace*{10pt}


\thispagestyle{empty}
\pagebreak

\tableofcontents
\thispagestyle{empty}

\pagebreak

\setcounter{page}{1}

\nocite{ysc2022}
\nocite{yscksh2022}
\nocite{kanikathottu2020aws}

\section{Introduction}

Over the past few years, many businesses have started to use cloud services to operate their product. As this trend continues, security on the cloud is emerging as an important concept just as on-premise security. For instance, if a cloud architecture is insecure, there can be data breaches where clients' private information is leaked. Also, the loss of assets such as Kubernetes clusters or computing instances will affect service availability. Moreover in Korea, there are certain certifications based on the Personal Information Protection Act such as ISMS or ISMS-P \cite{ismsp2022} that require an IT service to implement the required security functionalities. If the service doesn't satisfy the security requirements, the certification is rejected and the service may be penalized heavily. Thus, cloud security incidents may result in the same catastrophic events as on-premise security breaches.

Unfortunately, small companies or lean startups often do not have enough resources to design such secure architectures on the cloud. Consortium of CERT (CONCERT) recognized that this problem must be dealt with in the future. In the Hacking and Defense Contest (HDCON) 2021, they gave out a problem about designing a secure cloud architecture for small companies. \cite{hdcon2021} The problem describes a hypothetical IT company that uses both cloud services and on-premise machines. The chief privacy officer (CPO) knows that their current cloud infrastructure does not meet the required security standards, and wants to improve their architecture. However, the developers are fairly new to the cloud service and are having a hard time figuring out how things work on the cloud. Regarding this situation, the contestants were required to design and suggest a secure cloud architecture appropriate for this company. The problem modeled by CONCERT was very realistic, considering that the problem reflects the reality of small companies where developers are not very familiar with the cloud service itself or have no clue about how to improve the security on the cloud.

Thus for small organizations, it is of great importance to design a secure cloud architecture that can be deployed efficiently. Since access management is easy to apply and a key to enhancing overall security, we would like to propose a method concerning access management. Our main observation is that the separation of assets in multiple accounts results in simplified access control since accesses are automatically restricted within an account. In this paper, we introduce an effective strategy for designing a secure cloud architecture on AWS by using multiple accounts. We compare this strategy with the single account structure and explore the additional benefits of multi-accounts from the standpoint of the AWS Well-Architected Framework. Furthermore, we provide supplementary methods such as policy management and security auditing that help achieve operational excellence under the multi-account structure.

We restrict our discussion to the Amazon Web Service (AWS) since it is one of the most widely used cloud service providers. We also note that this discussion does not lose generality since analogous constructions are possible in other cloud services as well.

\section{Background}

\subsection{AWS Well-Architected Framework}

AWS provides a guide called the \textbf{AWS Well-Architected Framework} for building secure, stable, and flexible cloud infrastructures. \cite{aws:well-architected-framework} The framework is written by AWS solutions architects who have years of experience and covers a variety of use cases and best practices that can be efficiently adapted for many different organizations. Although the Well-Architected Framework is not a gold standard, it informs us of the general considerations for designing architectures. Therefore AWS advises that the framework should be adapted appropriately depending on the current cloud infrastructure. Specifically, AWS provides a service called the AWS Well-Architected Tool, which evaluates the current architecture and points out what can be improved from the current state. \cite{aws:well-architected-tool}

The framework consists of six pillars: operational excellence, \textbf{security}, reliability, performance efficiency, cost optimization, and sustainability.
Our main consideration will be the pillar of security, but we emphasize that these pillars are not independent of each other. In fact, they are complementary. For example, improving an architecture based on the pillar of security not only enhances security but also results in enhancements in other pillars as well. We will show that using the multi-account strategy enhances pillars of operational excellence, security, reliability, and cost optimization.

\subsection{The Five Pillars of Security}

The ultimate goal of the pillar of security is to secure the cloud infrastructure so that the business can be provided reliably without any interruption in a safe environment. To achieve this goal, the pillar of security aims to protect all assets on the cloud and prepare for unexpected attacks. The five pillars of security show how this is done in detail.

\begin{itemize}
    \item \textbf{Identity and Access Management (IAM)} aims to build a safe cloud environment where access to cloud services and resources is managed securely.

    \item \textbf{Detection} aims to build a system that collects and monitors logs from many cloud services and resources in use, to audit the system or detect anomalies.

    \item \textbf{Data Protection} aims to protect the data in the cloud, both in transit and at rest.

    \item \textbf{Infrastructure Security} aims to analyze threats and attack surfaces on the cloud, to set up appropriate defense mechanisms in advance.

    \item \textbf{Incident Response} aims to prepare for actual attacks by establishing and documenting procedures that can be used to mitigate attacks.
\end{itemize}

\subsection{Identity \& Access Management (IAM)}

Identity and Access Management (IAM) is a pillar of security concerning access to resources on the AWS cloud. In AWS, a service with the same name exists exactly for this purpose. \cite{aws:IAM} Using the IAM service, users can efficiently manage the authentication of identities and the authorization of actions.

\subsubsection{Identities and Actions}

\textbf{Identities} can be anything that can read or modify assets on the cloud. Usually, they are users who log in to the AWS web console or execute commands in the CLI. Sometimes, they can be programs that have a security credential, such as an application that is given an access key for making requests to the AWS cloud. Since the authentication part is done automatically by AWS, the administrator's job will be to manage user credentials and security credentials so that this information is not leaked outside.

\textbf{Actions} describe what can be done by identities on the cloud. They mainly consist of creating, reading, updating, and deleting (CRUD) resources. Here are some examples of actions on AWS: creating an EC2 instance, reading the contents of an EBS volume, updating the contents of an S3 bucket, and deleting an SSL certificate. Since these actions directly manipulate the resources on the cloud, it is of great importance to restrict the possible actions that can be performed on a resource. The IAM service enables the user to carefully customize the action allowed for an identity.

\subsubsection{IAM Policies}\label{sec:iam-policies}

In the IAM service, the administrator defines the kind of actions that an identity can or cannot perform. These access configurations are collected and represented as a statement inside a \textbf{policy}, as shown in \cref{lst:policy}.

\begin{figure}[t]
    \captionsetup{font=small}
    \begin{lstlisting}[
        language=json, captionpos=b, label=lst:policy,
        caption=An example policy allowing any actions on a DynamoDB table named \tt{Books}.,
    ]
{
  "Version": "2012-10-17",
  "Statement": [
    {
      "Effect": "Allow",
      "Action": "dynamodb:*",
      "Resource": "arn:aws:dynamodb:ap-northeast-2:123456789012:table/Books"
    }
  ]
}
    \end{lstlisting}
\end{figure}

Each statement consists of the \tt{Effect}, \tt{Action}, \tt{Resource}, and \tt{Condition} fields. The \tt{Effect} field is either \tt{Allow} or \tt{Deny}, describing whether to authorize this action. The \tt{Action} field is the action defined above, describing the kind of operations that can be done. The \tt{Resource} and \tt{Condition} fields enable detailed customization. If a \tt{Resource} value is set, the action only affects the specified resource. Lastly, the \tt{Condition} field is evaluated whenever an action is performed, and the statement is taken into consideration only if the condition is true. This customization will be helpful when we adjust policies using the principle of least privilege in \cref{subsec:principle-of-least-privilege}.

Users can define custom policies or use policies predefined by AWS. These policies are attached to an identity, and then the policy will take effect. If no policies are attached to an identity, all actions are implicitly denied by default, so attaching a policy is required to use AWS services.

Using the mechanisms described above, the administrator can effectively manage identities and their allowed actions. This is of great significance in cloud security since the system is insecure without any access control. Thus, IAM is the most important pillar among the five pillars of security, and we strongly suggest that one starts with defining IAM properly when designing a new cloud architecture.

\pagebreak

\section{The Multi-Account Strategy}\label{sec:multi-account-strategy}

Now we introduce the multi-account strategy. As the name suggests, the general strategy is to use multiple accounts and separate the assets in different accounts.

In this section, we elaborate on the drawbacks of single-account structures and show how multi-account structures solve them. We also introduce the AWS IAM Identity Center, which provides a single sign-on feature for managing multiple accounts efficiently. Lastly, we show an example of a multi-account structure that can be flexibly adapted for many different organizational structures.

\subsection{Drawbacks of Single Account Structures}

\begin{figure}[t]
    \centering
    \captionsetup{font=small}
    \includegraphics[width=0.8\textwidth]{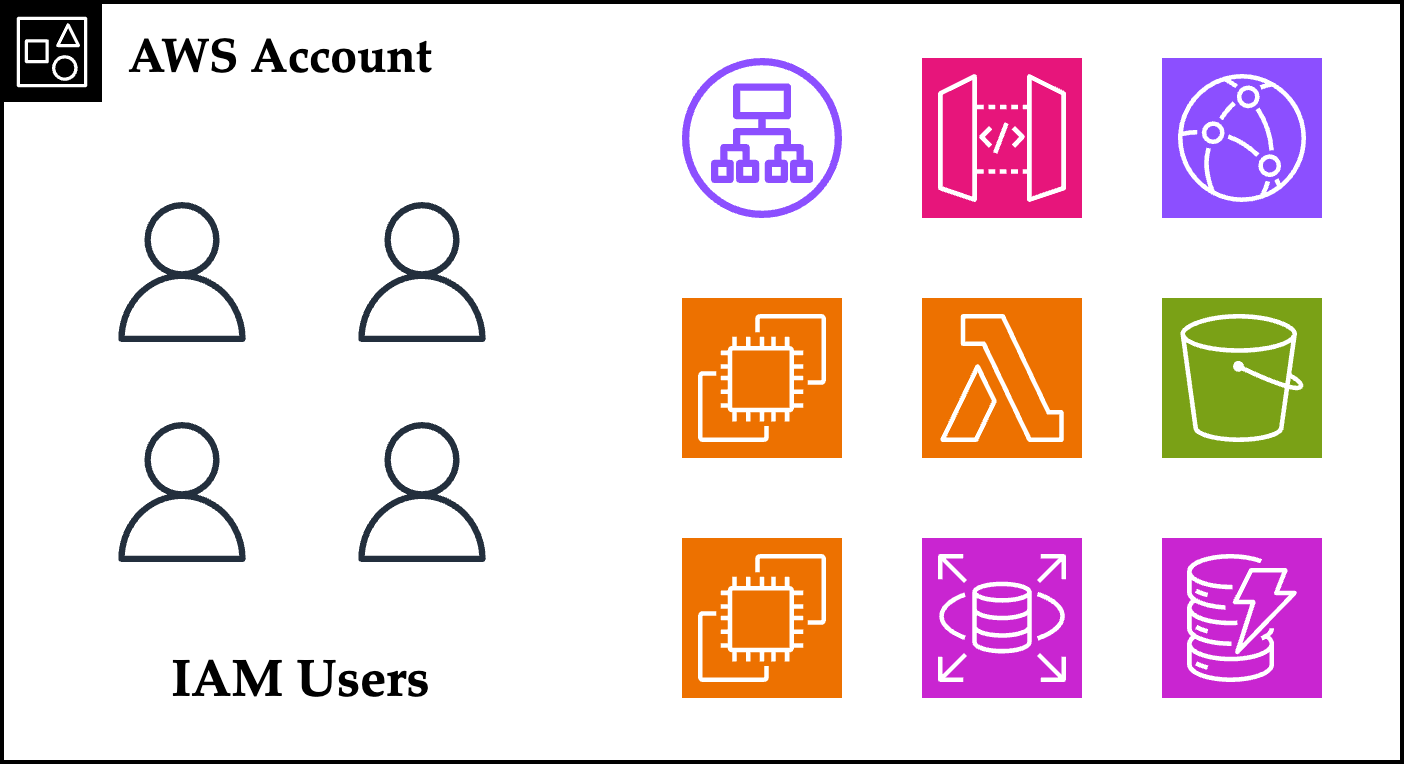}
    \caption{In single-account structures, all cloud resources are put inside a single account and all users access the same account. This may be efficient for small organizations, but it is insecure due to problems in visibility, environment separation and cost optimization.}
    \label{fig:single-account}
\end{figure}

In a small startup, where there are only a few developers, it may be efficient to use a single AWS account and create multiple IAM users under that account, as shown in \cref{fig:single-account}. This single account structure works for a small team, but managing policies gets tedious as more developers join. Suppose a hypothetical company where there are many teams, many different products, and about a hundred developers in total. If this company were to use the single account structure, here are some potential drawbacks of this structure.

First of all, there is the visibility problem. All resources in this account are visible to all developers of the company. Assume that each product uses EC2 instances for servers, S3 for static files, and RDS for the database. Then, all these assets for respective products will be accessible to any developer, which is insecure. Especially, databases often contain the private data of customers, thus no access control on the databases is against privacy.

Additionally, there may be multiple instances of EC2 and RDS databases for horizontal scaling. Then the developers of each product will have a hard time distinguishing which instance is theirs. This increases the risk of confusion or mistakes such as accidental deletion.

To resolve this problem, we could use fine-grained policies that make use of the \tt{Resource} and \tt{Condition} fields. By using those fields, the administrator could define policies that only allow actions for a specific set of resources. However, this method lacks both flexibility and scalability since the policies are too specific. The administrator must inspect all resources in the account and group them by considering which teams or products use that resource. Then policies must be manually created for each team or product, and manually attached to each developer. Even for a small change in the resources or the product architecture, the administrator may end up modifying a lot of statements in a policy. This wouldn't be an ideal way to manage policies.

Secondly, environment separation is hard in a single account. When developing software, it is customary to set up a development server which is used for internal testing purposes. Now suppose that the development server is set up in the same account as the production server. Since there are many products, this results in multiple environments spanning multiple products. In this situation, any problem in the cloud can potentially have a large area of influence, which is undesirable. For example, a security breach, an error in the cloud, or maybe a developer's mistake can affect multiple environments and products since they are not separated. As another example, if the development and production servers share the same VPC network, network-related tests may affect the production server. Performing stress tests on the development server will use up the network bandwidth and affect the production server, slowing down connections of users of the product.

Lastly, there is a cost optimization problem. Without any additional setup, it is difficult to measure the exact amount of costs spent by a specific team. Then during cost analysis, it will be hard to detect what costs can be reduced, and what resource is being charged for. As a result, it is hard to figure out possible optimizations, and some unused resources may be left unnoticed for a long time, causing additional charges. Although this functionality can be achieved by using cost allocation tags, all resources must have a tag for this to work, which is cumbersome. \cite{aws:cost-allocation-tags}

\subsection{Benefits of Multi-Account Structures}

The problems of single account structures can be easily solved by using multiple accounts, basically because separation of cloud assets is inevitable when multiple accounts are used.

The visibility problem is easily solved by creating a separate account for each group, whether a team or a product. If each group uses a separate account, only the developers of that group can access the resources inside that account, and other developers are not given access. So, developers can access a resource if and only if it is their group's resource, which is secure. Also, compared to single accounts where the administrator had to manually create fine-grained policies, the access control is now done automatically by the separation of accounts. Additionally, fewer developers use the same account, so fine-grained access control is much more manageable.

\begin{figure}[t]
    \centering
    \captionsetup{font=small}
    \includegraphics[width=\textwidth]{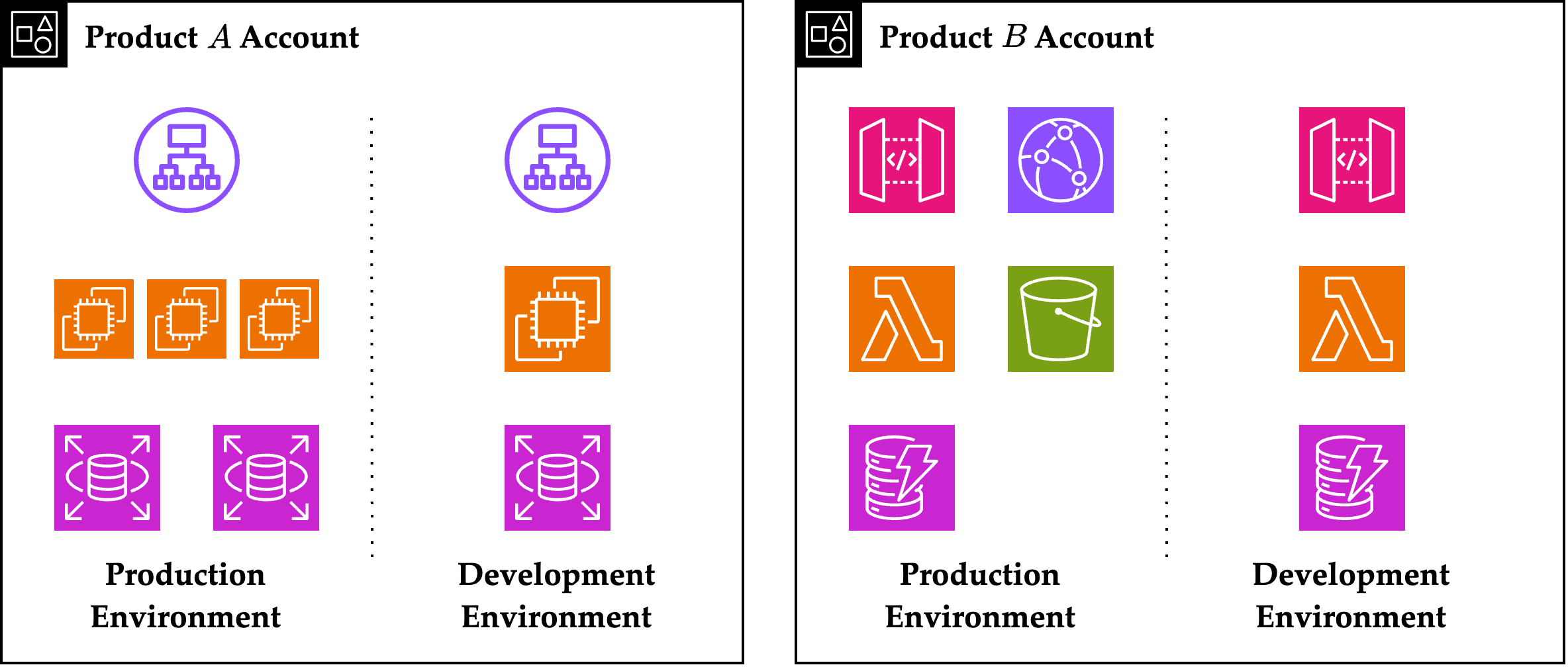}
    \caption{Separation by product, where each product uses a dedicated account.}
    \label{fig:separation-by-product}
\end{figure}

For the remaining two problems, environments can also be separated easily using multiple accounts. By creating a different account for each environment, separation is easily achieved. The area of influence on incidents is greatly reduced to a single account, only affecting the resources in it. Finally, the cost optimization problem is solved since accounts are charged separately, and the usage of resources will be easier to analyze inside a single account.

Besides these benefits, the multi-account structure is scalable and flexible. For example, when the company is launching a new product or creating new teams, a new account is easily created for use. Also, member changes in a team can be handled easily. When a developer moves to a different team, the IAM User from the previous team's account is deleted, and a new one is created in the current team's account.

To summarize, these benefits of a multi-account structure fall into the category of four pillars from the AWS Well-Architected Framework. The first one is the pillar of operational excellence. The company's AWS administrator now has more control over each account, since the scope of each account has been reduced. Management of policies and assets in each account is much easier than managing everything inside a single account, resulting in efficient operations on the cloud infrastructure. The second one is the pillar of reliability. Separation of environments leads to reduced area of influence, increasing the reliability of an application. The third one is the pillar of cost optimization, where costs can be analyzed and optimized effectively by inspecting each account. The last one is the pillar of security, which we discuss in detail using the five pillars of security.

\begin{itemize}
    \item \textbf{Identity and Access Management}: we gain better control of access to resources since accounts automatically restrict the access boundaries.
    \item \textbf{Detection}: it is easier to audit smaller accounts, and we can easily detect errors in them.
    \item \textbf{Data Protection}: privacy is enhanced since developers can only access the data of their products.
    \item \textbf{Infrastructure Security}: analyzing threats for each account is easier, and the attack surface is reduced by separating accounts.
    \item \textbf{Incident Response}: it is easier to prepare for attacks on separate accounts, and appropriate procedures can be prepared inside each account.
\end{itemize}

\subsection{Configuring Multi-Account Structures}

As discussed, the multi-account structure enhances the overall security of the cloud. However, the usage of multiple accounts also gives rise to new problems. Thus, additional configuration is required to fully utilize the multi-account structure. We describe two scenarios where this is required and explain how to handle them using the mechanics of AWS.

\subsubsection{Handling the Overhead of Multiple Accounts}\label{subsubsec:overhead-multiple-accounts}

First of all, it is difficult to provision and manage multiple accounts. To sign up for an AWS account, lots of information including user details and billing must be filled in. Moreover, verification with SMS or phone is required, complicating the whole process. This must be done for each account, which easily becomes complex. Also, multiple accounts have to be managed just like teams and employees are managed in a company. But the accounts are currently unstructured, and unrelated to each other, introducing management overhead.

Suppose we went through to provision all accounts. Now that we have multiple accounts, the cloud administrator has to attach policies for individual developers, which is a lot of repeated work. Also, the same policy is repeatedly used in different accounts because developers in the same position are likely to be given the same access privileges. For example, backend developers often have access to EC2 or RDS services, and frontend developers often access S3 or CloudFront. Due to redundancy, it would be frustrating for the administrator to create the same policies for multiple accounts.

Another critical problem is that switching between accounts is bothersome. This involves developers involved in many projects or groups and developers who have access to multiple environments in different accounts. During work, a developer may have to switch accounts back and forth for various reasons, such as testing on the development environment and then deploying it to production. To switch accounts, one must log in again, and multi-factor authentication will be required every single time. Furthermore, one can't keep two login sessions in a single browser. If there is a different login session in another tab, that session will automatically expire on a new login. To get around this, developers are forced to use multiple profiles or incognito tabs in the browser.

Fortunately, AWS offers a service called \textbf{AWS IAM Identity Center} that solves the above problems by using the single sign-on (SSO) feature. This service provides the core functionalities required for the effective management of multiple accounts, and this is a must-have for many organizations. Considering its importance, we discuss it in detail in \cref{subsec:iam-identity-center}.

\subsubsection{Sharing Resources Between Accounts}

Depending on the organization or product, some resources need to be shared between multiple accounts. We describe two common examples of cross-account resource sharing, static files in S3 buckets and container images in the Elastic Container Registry (ECR).

Static files are often stored inside S3 buckets, so replicating the whole bucket in multiple accounts is not necessary. Also, replication is expensive if the files are large, such as AI model parameters and datasets that need to be shared. Instead, we can keep a single source bucket and configure it so that multiple accounts can fetch data from the bucket. Similarly for container images in ECR, we only need to keep a single registry for each kind of image. It is hard to manage different versions of the same image spread out in multiple accounts, so it is better to centralize it. So we must configure the registry so that multiple accounts can pull images from a centralized container registry.

\begin{figure}[t]
    \centering
    \captionsetup{font=small}
    \includegraphics[width=\textwidth]{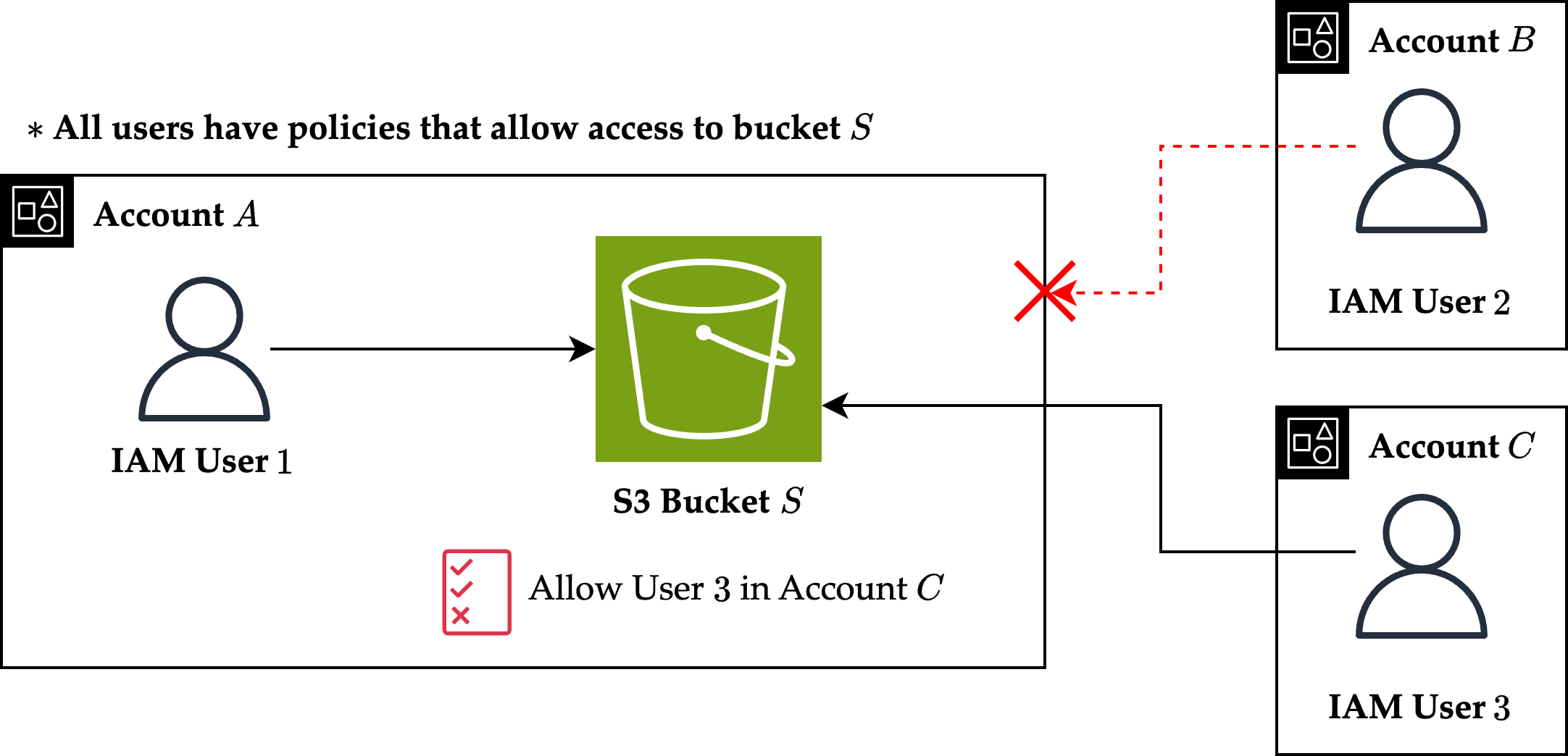}
    \caption{Bucket \(S\) explicitly allows access for user \(3\), but implicitly denies user \(2\). Both identity-based and resource-based policies must explicitly allow the access for cross-account accesses to work.}
    \label{fig:resource-based-policy}
\end{figure}

These cross-account resource sharing scenarios can be handled by \textbf{resource-based policies}. \cite{aws:resource-based-policy} Unlike the identity-based policies described in \cref{sec:iam-policies} which are attached to identities, resource-based policies are attached to resources to achieve additional access control over a resource. Essentially, resource-based policies configure the identities that can access a resource. These policies are rarely used in single account settings, but they are useful when it comes to sharing resources. \Cref{fig:resource-based-policy} shows why this is the case.

We assume that all users are already given identity-based policies that allow access to bucket \(S\) and that the bucket has a resource-based policy allowing access for user \(3\) in account \(C\). Recall that the default behavior of policies is implicit denial. But user \(1\) can access \(S\) even though \(S\) does not explicitly allow it. This is because user \(1\) is in the same account as \(S\). Thus, additional configuration is not required for users in the same account, which explains why resource-based policies are often unused in single-account architectures.

On the other hand, the situation is different for users in other accounts. For cross-account accesses, both identity-based and resource-based policies must explicitly allow access to the resource. Consequently, user \(2\) is denied, but user \(3\) is allowed access to \(S\). Thus, additional work of attaching resource-based policies is required to enable cross-account accesses. As shown in \cref{fig:resource-based-policy}, the bucket's resource-based policy must be modified.

Another way to share resources is to use the AWS Resource Access Manager (RAM). \cite{aws:ram} AWS RAM service is especially useful for cases where resource-based policies are not applicable. A hosted zone in Amazon Route 53 DNS service is an example. Since different environments under the same product often share the same hosted zone, sharing DNS records would allow modification of records from both accounts, which is convenient. In this case, the RAM service can be utilized to share a hosted zone between different environments under the same product.

\subsection{Single Sign-On with AWS IAM Identity Center}\label{subsec:iam-identity-center}

As stated in \cref{subsubsec:overhead-multiple-accounts}, using multiple accounts can introduce additional overhead for management. Thankfully, we can use the features of \textbf{AWS IAM Identity Center} to manage multiple accounts efficiently. \cite{aws:identity-center}

\subsubsection{Account Structuring with AWS Organizations}

To use the IAM Identity Center, one must first enable the organization feature in the \textbf{AWS Organizations} service. \cite{aws:organizations} By using AWS Organizations, accounts can be managed similarly as employees or teams are managed in a company. Specifically, the AWS Organizations service supports organizational structure through organizational units.

An \textbf{organizational unit} (OU) is a group of accounts that can be managed as if they were a single unit. As an analogy, an account in an organizational unit is just like an employee who belongs to a team in a company. Thus, organizational units can be used to group accounts together, to form an organizational structure. The structure gives a high-level overview of accounts and is helpful for the administrator since it allows the management of groups of accounts, rather than individual accounts.

The account that enabled AWS Organizations becomes the \textbf{management account} of the organization, and operations on organization units or provisioning of new accounts must be done from this account. Also, provisioning an account is simple using AWS Organizations. All we need is the account name and the email, and the provisioned account will automatically join the organization.

\subsubsection{SSO Users and Groups}

After enabling AWS Organizations, we can use the AWS IAM Identity Center. Essentially, this service provides a \textbf{single sign-on} (SSO) feature for multiple accounts in the organization.\footnote{This service was formerly known as the AWS Single Sign-On.} With the SSO feature, developers only have to sign in once to the SSO console, and switching between accounts does not require additional logins, which is convenient.

\begin{figure}[t]
    \centering
    \captionsetup{font=small}
    \includegraphics[width=\textwidth]{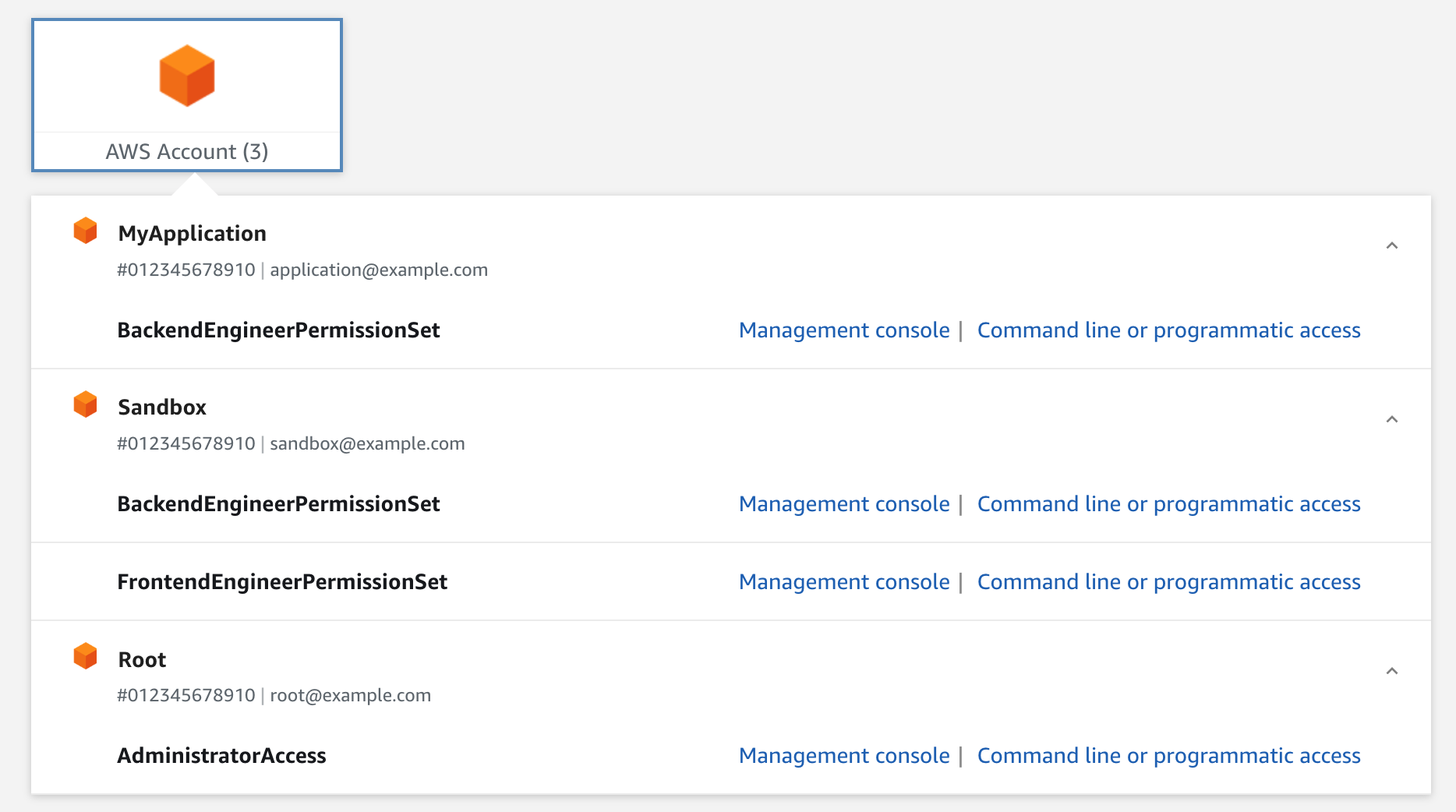}
    \caption{The AWS Single Sign-On console shows a list of accounts and the permission sets that can be used to access that account. This SSO user can access the \tt{MyApplication} account with the permissions assigned to a backend engineer. The console also enables users to switch accounts easily.}
    \label{fig:sso-console}
\end{figure}

To use the SSO feature, developers must be added as \textbf{SSO users}. The definition is obvious from the name, but there is a critical difference between SSO users and IAM Users. IAM Users are bound to a single account but SSO users are not, since SSO users can log in to multiple accounts. By using SSO users, IAM Users are not needed in each account. This implies that developers with access to multiple accounts only have to be added once as an SSO user, simplifying user management.

Additionally, SSO users can be organized into groups, which also allows the management of multiple users as a single unit. This will simplify the process when policies are attached to groups, instead of individual SSO users.

\subsubsection{Permission Sets}

Now that we have accounts and SSO users/groups prepared, policies have to be created in each account by the administrator. The problem was that the same policy had to be created repeatedly in each account. However, if we use permission sets, we can configure policies that can be used for multiple accounts.

A \textbf{permission set} is a collection of policies that can be reused for different accounts, SSO users, and groups. In contrast to policies that were attached to IAM Users within an account, a permission set is assigned to a user-account pair or a group-account pair. If a permission set is assigned to a user-account pair, then the SSO user can access the account with the policies defined in the permission set. Similarly, for a group-account pair, every user in the group can access the account with the specified policies.

Considering that developers in the same position tend to require similar policies, a permission set can be created specifically for that position, and can be used to assign multiple users to multiple accounts. For instance, we can create a permission set for backend developers, allowing access to EC2 instances and RDS databases. Then this permission set can be assigned multiple times on backend developers (or groups) over different product accounts. This feature optimizes policy management, allowing efficient operation in multiple account settings.

\subsection{An Example of Multi-Account Structure}

We dedicate this section to show an example of a multi-account structure, including organizational units and permission sets. However, our example is not a definite answer since there are many ways to design an account structure depending on the specific use case. Furthermore, architectures can always change over time. Thus, the example serves as a purpose to suggest what design choices are possible, and what factors should be considered before designing a multi-account structure.

Consider a hypothetical company with three teams: Red, Blue, and internal team. The developers of Team Red are working on product \(A\) which is a large-scale service with a lot of active users. Team Blue works on two products \(B\) and \(C\). Product \(B\) is their main product, which they are actively developing, and \(C\) is a small web service not actively being maintained. We assume that teams Red and Blue both have frontend and backend developers. Lastly, the internal team consists of data analysts and full-stack developers who work on internally used products such as testing tools or Slackbots. This company originally used a single-account structure, but now wants to migrate to a multi-account structure.

\subsubsection{Organizations and Account Structure}

\begin{figure}[t]
    \centering
    \captionsetup{font=small}
    \includegraphics[width=\textwidth]{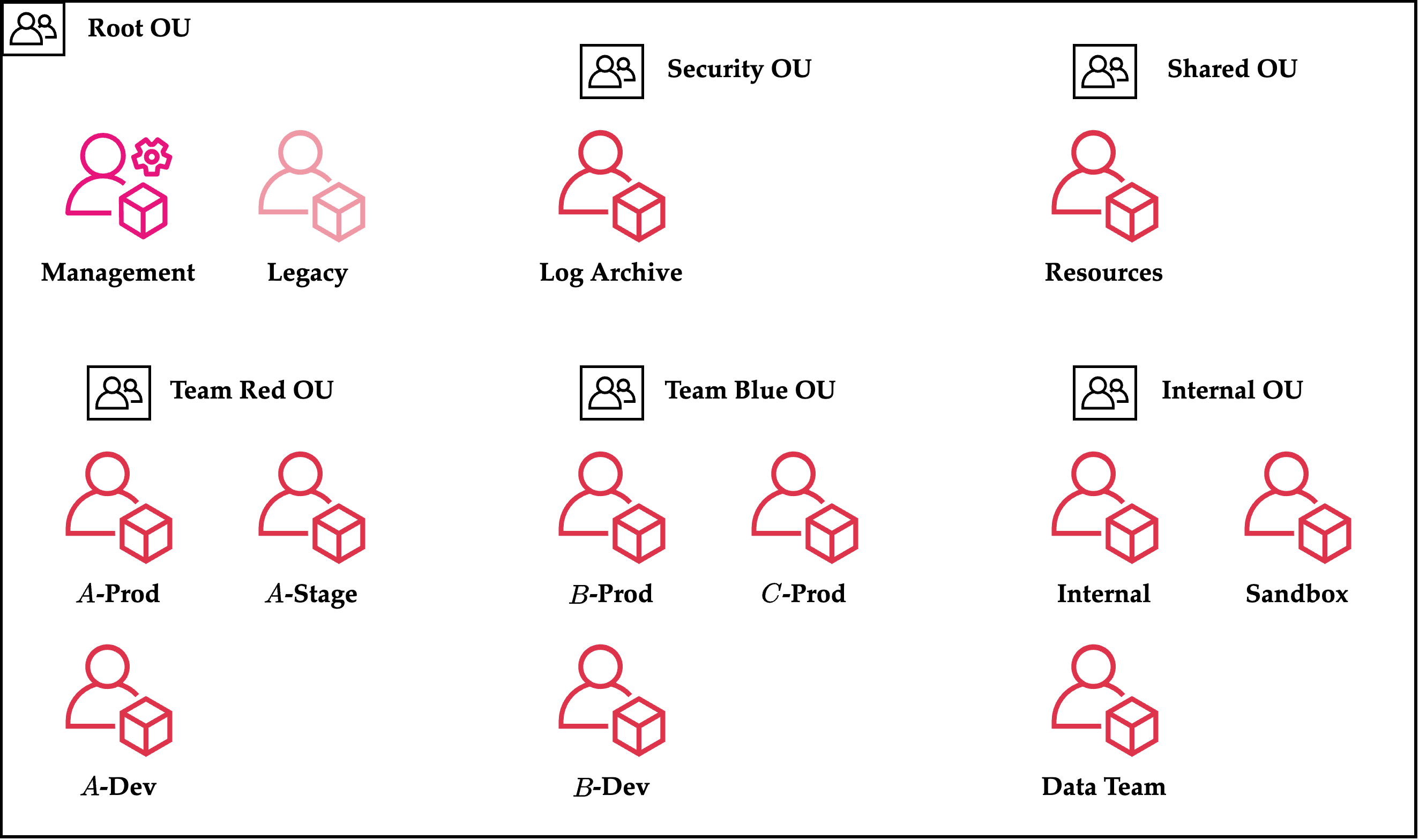}
    \caption{An example of a multi-account structure for a hypothetical company. Accounts are structured into five organizational units, and accounts are provisioned under each unit.}
    \label{fig:multi-account-example}
\end{figure}

An example structure is given in \cref{fig:multi-account-example}. At the highest level, there is the root organizational unit, and the remaining organizational units are on the second level. Each team gets its organizational unit, and two additional organizational units were created for shared resources and security purposes.

Now we explain the possible reasons for choosing this design. First of all, the legacy account is left for the case where migration of cloud resources is incomplete. This account will be removed when it is no longer required.

As for the accounts of Team Red, we suggest using \(3\) accounts, each dedicated to a separate environment. Separating the development and production environments is acceptable, but there is another account for the staging environment. The reasoning behind this is that product \(A\) is a large-scale service, that may require a lot of testing before being deployed in production. However, if the team chooses not to use a staging environment or decides to keep the staging environment with the development environment, the staging account can be deleted afterwards.

Team Blue had two products in operation. We suggest using \(2\) accounts for product \(B\), each for an environment. As for product \(C\), it is a small service that is not actively maintained, so a single account suffices, and separating accounts for each environment is not required. This is because too much separation may lead to higher overhead for small services.

Next for the internal team, the data analysts are given a single account, and the internal product developers are also given a single account. In these cases, strict separation is not necessarily required, since these resources are only used internally. If the internal team grows large, additional accounts can be provisioned to separate each internal product. The sandbox account is an account that can be used for any purpose. They are useful for experimenting with new architectures, testing new features of AWS, and more. It is also used internally, so a single account suffices.

For the last two organizational units, they are not strictly required but it helps to have them. The resources account in the shared unit is a storage of shared resources, that can be accessed by other accounts. Rather than putting a shared resource in a product account, it is better to have an account dedicated to shared resources. Lastly, the log archive account in the security unit is an account for collecting service logs from all other accounts. The logs are collected and stored in a single account, which can be convenient for auditing purposes.

As seen from the example, accounts can be split with different criteria. Accounts for product \(A\) is a case where accounts are split by environments, having separate accounts for product \(A\) and \(B\) is splitting by product. Also, a whole team can use a single account as in the data team account or internal account. Lastly, accounts can be separated for dedicated purposes, just as in the sandbox or resources account. But one should be careful since too much separation can lead to more overhead. In summary, account structures should be designed in a way that best fits the organizational structure, and during the design process, factors like team size, environment, and products should be carefully considered.

\subsubsection{Structure of Permission Sets}

\begin{figure}[t]
    \centering
    \captionsetup{font=small}
    \includegraphics[width=\textwidth]{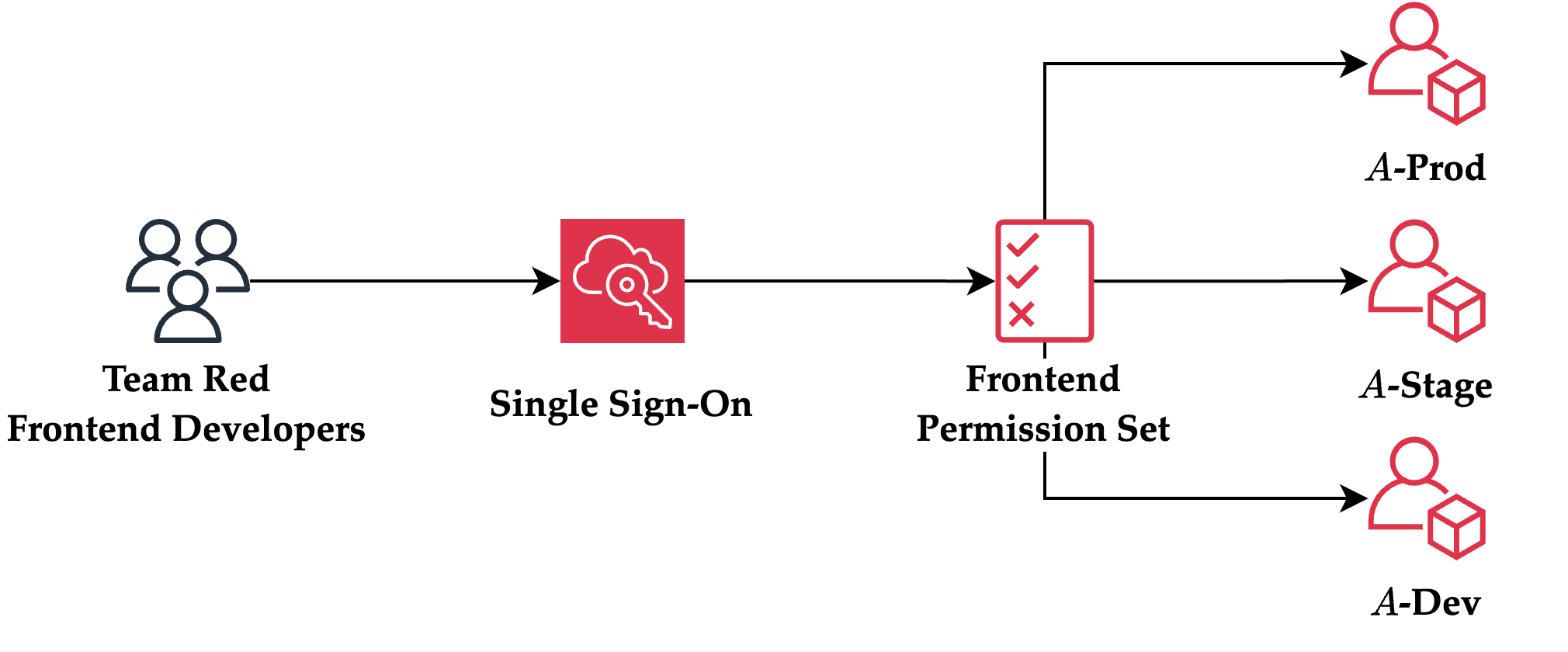}
    \caption{Permission sets can be assigned multiple times, reducing redundancies. The frontend permission set is assigned to the frontend developer group of Team Red and the accounts of product \(A\).}
    \label{fig:frontend-permission-set}
\end{figure}

Teams Red and Blue both consist of backend and frontend developers, so we suggest creating a permission set for each developer position. Then the backend permission set should be assigned so that the backend developers of product \(A\) can use it to access accounts of product \(A\). The same goes for the frontend permission set, for products \(B\) and \(C\). The same policy can be reused with different environments since developers will require similar policies regardless of the environment. As for the internal account, the developers are full-stack, so create a separate permission set appropriate for full-stack development. The sandbox account is for experimenting, so we create a permission set with full access to all resources and let any developer use it. Continuing in this fashion, we can define a policy for each group or purpose and attach it to multiple group-account pairs, eliminating redundancy.

However, there are some special cases to be handled. For team lead developers, it may be convenient to give them full access to all resources for accounts on their team. As for shared resources in the resources account, cross-account access is required, so resource-based policies must be set for individual resources, and identity-based policies must be included in the permission sets.

In summary, to design permission sets, a starting point would be: considering the policies required for each position of developers, or the policies required for maintaining each product. After that, special cases like team leaders and policies for cross-account accesses should be handled.

\section{Secure Policy Management in Multiple Accounts}

\nocite{aws:iam-best-practices}

In the previous section, we explained the multi-account structure in detail and how to design such structures that use the features of AWS IAM Identity Center. We discussed how to reduce redundancy of policies by using permission sets, and thus policies are centralized as permission sets in the management account of the organization. Now that the number of permission sets is greatly reduced, we turn our attention to the actual content of the policies in those permission sets.

\subsection{The Principle of Least Privilege}\label{subsec:principle-of-least-privilege}

For security, any policy should not allow access that isn't necessarily required. The principle of least privilege deals with this and alerts us to manage the contents of policies securely so that no users are granted too much privileges.

\textbf{The principle of least privilege} states that no identity should be given more access privileges than required. This is the most important principle in identity and access management, but also the most difficult one to achieve. Ideally, a user must only have the necessary access privileges, but access privileges keep changing depending on time or the roles and responsibilities of the user. Therefore, we cannot force every single user to adhere to the principle. If we always kept privileges at a minimum, the security administrator would be overflooded by temporary access requests. This is inefficient for both users and administrators, so we relax the conditions instead. Hence, our objective is to keep access privileges minimal, but still allow a little freedom so that users can perform their tasks without frequently requesting temporary access privileges.

\subsection{Levels of IAM Actions}

Recall that IAM actions specify what can be done by identities on the cloud. They are written in the form of \tt{service:operation}, denoting the kind of operation in some service. For example, the action \tt{s3:GetObject} is the read operation in the S3 service. The syntax also supports wildcards (\(\ast\)), allowing actions such as \(\tt{s3:Put}\ast\), which includes all operations that start with \tt{Put} such as \tt{PutObject}. Then this action describes update operations in the S3 service. Using this syntax, IAM actions can be organized into four levels of hierarchies.
\begin{enumerate}
    \item Full access \(*\tt{:}*\), which describes any operation in any service.
    \item Full access within a specific service \(\tt{service:}*\), such as \(\tt{s3:}*\).
    \item Read-only or write-only access within a specific service, such as \(\tt{s3:Get}*\) or \(\tt{s3:Put}*\).
    \item A specific action, such as \tt{s3:PutObject}.
\end{enumerate}

This hierarchy is also reflected in the default AWS-managed policies. \cite{aws:aws-managed-policies} The policy \tt{Admin\-istratorAccess} contains level \(1\) action and policies like \tt{AmazonEC2FullAccess} or \tt{CloudFront\-FullAccess} include level \(2\) actions. Lastly, level \(3\) actions can be found in policies such as \tt{AmazonS3ReadOnlyAccess} and \tt{IAMReadOnlyAccess}.

\subsection{Setting Appropriate Actions for Least Privilege Permissions}

All policies must consist of level \(4\) actions to achieve the principle of least privilege, but this is impossible. To use level \(4\) actions, security administrators must first inspect the kinds of tasks each developer performs on AWS. Then they have to list all the specific actions required for each developer, which is very time-consuming. The AWS system is so complex that a click on the console may require multiple actions to be allowed. Even with the AWS official documentation, it will be hard to find all the required actions, without omitting any.

Besides, if all policies consisted of level \(4\) actions, all actions not included in the policy will be implicitly denied. Since roles and responsibilities change over time, developers are sure to encounter a situation where additional access privileges are required. Developers have no choice but to make an access request to the administrator, and administrators will be flooded with requests. Moreover, it doesn't end at granting requests, they also have to be revoked according to the principle if they aren't being used. This is too much burden on the security administrators. Thus, level \(2\) to \(3\) actions are practical for usage.

It is hard to find the required level \(2\) or \(3\) actions from the beginning, so we suggest an incremental approach. First, list the services that a team or position uses, and start with level \(2\) actions of those services. Then gradually switch to level \(3\) by narrowing down the actions that are required. After that, further customization with fine-grained policies is also possible, using the \tt{Resource} or \tt{Condition} fields in policy statements.

As an example, consider the permissions of a backend developer. Backend developers often use EC2 instances, certificate managers (ACM) for setting up HTTPS, RDS databases, and sometimes use IAM service for setting instance roles. Including all these, a backend developer permission set can be initially configured to contain full access to the mentioned services. As developers use AWS, there will be unused actions such as creating new IAM resources or new certificates. In this case, the actions can be edited so that full access to ACM and IAM are restricted to read-only access.

\subsubsection{AWS IAM Access Analyzer}

AWS offers the \textbf{IAM Access Analyzer} for finding the least privileged permission of a user. \cites{aws:iam-access-analyzer} The access analyzer shows the last used date for each policy, so it can be used to find unused policies or policies that haven't been used for a long time. With this feature, the administrator can adjust the policies inside permission sets according to the principle of least privilege. Thus the periodic usage of the access analyzer is strongly recommended, for revoking unused policies. Additionally, the access analyzer collects information from the user's recent activities on the cloud and even generates a least privileged policy for that user. \cite{aws:access-analyzer-policy-generation} The generated policy might be too strict, so it can be modified further to introduce a little flexibility.

\section{Auditing Multiple Accounts}

It is important to keep multiple accounts compliant with the security policies of the organization. In this section, we introduce methods to audit multiple accounts and secure them.

\subsection{Importance of Documentation}

After designing the multi-account structure and setting up permissions, the final architecture must be documented in detail, and its revision history should be kept for future reference. When the administrator audits the cloud infrastructure, these documents will be used as a reference for checking if the infrastructure is obeying the security policies, or for checking if there are any new changes. Especially, the permission sets or any important assets such as access credentials should be kept documented so that their changes are always recorded.

The documents also contribute to the overall security of the cloud. The documents will be read by the users, and they will acknowledge how the architecture was designed to improve security. Then the documents will serve as a security guide to the users, and they will be able to use the features of the cloud securely, without breaking the security rules.

However, we emphasize that these documents must be updated whenever there is a modification to the cloud infrastructure. If the documents are not updated accordingly or are long outdated, users will get confused which can lead to insecure operation on the cloud. Thus for both auditing and secure operations on the cloud, frequent documentation is important.

\subsection{Cloud Audit Logs}

For auditing cloud accounts, we can use cloud audit logs which contain the actions performed by all users in the cloud. In AWS, cloud audit logs are collected by \textbf{AWS CloudTrail}. \cite{aws:cloudtrail} It logs when the users logged in, what resources were created or deleted, and more. CloudTrail must be enabled for all accounts so that the audit logs can be used afterwards for detecting malicious activity or changes in the cloud infrastructure.

We also suggest using a centralized log archive account that collects the audit logs from all accounts of an organization, as previously shown in \cref{fig:multi-account-example}. Without a log account, tools for analyzing audit logs would have to be installed for every account in the organization. However, by using a centralized log archive, audit logs can be efficiently analyzed inside the account by creating alert rules or dashboards only in this account.

\section{Conclusion}

In summary, we illustrated the multi-account strategy in detail with examples and examined additional procedures for operational excellence in the multi-account structure. Identity and access management on the cloud is an extensive topic with endless potential enhancements for each organization. However, we emphasize that access management with the multi-account strategy is a great starting point for improving the overall security of the cloud, considering all the benefits it brings. We hope that many organizations adopt the multi-account strategy into their infrastructure and improve their security.

\pagebreak

\vspace*{10pt}

\phantomsection
\addcontentsline{toc}{section}{References}

\begin{center}
    {\fontsize{16}{0}\selectfont \textbf{References}}
\end{center}

\printbibliography[heading=none]

@misc{ysc2022,
    author      = {Sungchan Yi},
    title       = {개발자를 위한 AWS 클라우드 보안 (1) - 클라우드 설계 원칙과 IAM},
    url         = {https://tech.scatterlab.co.kr/aws-cloud-security-for-devs-1/},
    year        = {2022},
    month       = {5},
    urldate     = {2023-11-09}
}

@misc{yscksh2022,
    author      = {Sungchan Yi and Sunghun Kim},
    title       = {개발자를 위한 AWS 클라우드 보안 (2) - 로깅 및 모니터링과 데이터 보호},
    url         = {https://tech.scatterlab.co.kr/aws-cloud-security-for-devs-2/},
    year        = {2022},
    month       = {6},
    urldate     = {2023-11-13}

}

@book{kanikathottu2020aws,
    title={AWS Security Cookbook: Practical solutions for managing security policies, monitoring, auditing, and compliance with AWS},
    author={Kanikathottu, H.},
    isbn={9781838827427},
    year={2020},
    publisher={Packt Publishing}
}

@misc{hdcon2021,
    author      = {CONCERT},
    title       = {2021 해킹방어대회(HDCON) 참가 안내},
    url         = {https://www.concert.or.kr/bbs/board.php?bo_table=notice&wr_id=742},
    year        = {2021},
    month       = {10},
    urldate     = {2023-11-09}
}

@book{ismsp2022,
    author         = {KISA},
    title          = {정보보호 및 개인정보보호 관리체계 (ISMS-P) 인증기준 안내서},
    year           = {2022},
    month          = {4},
}

@misc{aws:well-architected-framework,
    title       = {AWS Well-Architected Framework},
    url         = {https://docs.aws.amazon.com/wellarchitected/latest/framework/welcome.html},
    year        = {2023},
    month       = {10},
    urldate     = {2023-11-13}
}

@misc{aws:well-architected-tool,
    url          = {https://aws.amazon.com/well-architected-tool/},
    title        = {AWS Well-Architected Tool},
    urldate      = {2023-11-13},
}

@misc{aws:IAM,
    url          = {https://aws.amazon.com/iam/},
    title        = {AWS Identity and Access Management},
    urldate      = {2023-11-16},
}

@misc{aws:cost-allocation-tags,
  url          = {https://docs.aws.amazon.com/awsaccountbilling/latest/aboutv2/cost-alloc-tags.html},
  title        = {Using AWS cost allocation tags},
  urldate      = {2023-11-14},
}

@misc{aws:resource-based-policy,
  url          = {https://docs.aws.amazon.com/IAM/latest/UserGuide/access_policies_identity-vs-resource.html},
  title        = {Identity-based policies and resource-based policies},
  urldate      = {2023-11-16}
}

@misc{aws:ram,
  url          = {https://aws.amazon.com/ram/},
  title        = {AWS Resource Access Manager},
  urldate      = {2023-11-16}
}

@misc{aws:identity-center,
  url          = {https://aws.amazon.com/iam/identity-center/},
  title        = {AWS IAM Identity Center},
  urldate      = {2023-11-16}
}

@misc{aws:organizations,
  url          = {https://aws.amazon.com/organizations/},
  title        = {AWS Organizations},
  urldate      = {2023-11-16}
}

@misc{aws:aws-managed-policies,
  url          = {https://docs.aws.amazon.com/aws-managed-policy/latest/reference/policy-list.html},
  title        = {AWS managed policies},
  urldate      = {2023-11-17}
}

@misc{aws:iam-best-practices,
  url          = {https://docs.aws.amazon.com/IAM/latest/UserGuide/best-practices.html},
  title        = {Security best practices in IAM},
  urldate      = {2023-11-17}
}

@misc{aws:iam-access-analyzer,
  url          = {https://docs.aws.amazon.com/IAM/latest/UserGuide/what-is-access-analyzer.html},
  title        = {Using AWS IAM Access Analyzer},
  urldate      = {2023-11-17}
}

@misc{aws:access-analyzer-policy-generation,
  url          = {https://docs.aws.amazon.com/IAM/latest/UserGuide/access-analyzer-policy-generation.html},
  title        = {IAM Access Analyer policy generation},
  urldate      = {2023-11-17}
}

@misc{aws:cloudtrail,
  url          = {https://aws.amazon.com/cloudtrail/},
  title        = {AWS CloudTrail},
  urldate      = {2023-11-17}
}

\pagebreak

\vspace*{8pt}

\begin{center}
    {\fontsize{16}{0}\selectfont \textbf{국문초록}}
\end{center}

\vspace*{10pt}

최근 많은 IT 기업이 편리하게 자사 제품을 배포하기 위해 클라우드 서비스를 사용한다. 이에 따라 기업의 클라우드 자원은 새로운 공격 표면이 되었고, 클라우드 보안이라는 분야가 새롭게 대두되었다. 그러나 클라우드 보안은 기존의 온프레미스 보안에 비해 충분히 강조되지 못해 보안에 취약한 클라우드 아키텍처를 사용하는 경우가 많다. 특히 작은 조직의 경우 안전한 클라우드 아키텍처를 고안할 인력이 부족한 경우가 많아 클라우드에서 발생하는 보안 사고에 취약한 편이다.

이 상황에서 보안을 손쉽게 강화하려면 다중 계정 환경을 적용하면 된다. 다중 계정 환경은 클라우드의 자원을 분리하고 관리 부하를 줄여 보안을 강화하는 전략으로, 노력 대비 큰 보안 향상을 준다. 이 전략을 적용하면 자동으로 접근 권한이 계정 범위 내로 제한되며, 정책 관리 시 발생하는 불필요한 중복이 제거된다. 안전한 아키텍처를 위해 권한 관리가 필수임을 고려한다면, 다중 계정 환경은 인력이 부족한 작은 조직에서도 적용할 수 있는 효과적인 보안 강화 방법이다.

이 논문에서는 다중 계정 환경의 장점을 단일 계정 환경과 비교하여 분석하고, AWS가 제공하는 서비스를 이용해 다중 계정 환경을 손쉽게 구성하는 방법을 설명한다. 또한 다중 계정 구조의 구체적인 예시를 통해 계정 구조 설계 시 유의할 점들을 언급한다. 마지막으로 최소 권한 원칙의 점진적 도입을 통한 안전한 정책 관리 방법과 다중 계정의 감사 방법을 소개하여 다중 계정 구조에서 운영 효율성을 달성하는 방법을 설명한다.

\vspace*{10pt}

\quad \textbf{주요어}: 다중 계정 환경, 권한 및 접근 제어, 클라우드 보안

\thispagestyle{empty}
\pagebreak

\end{document}